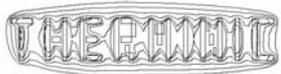

# Method of Images for the Fast Calculation of Temperature Distributions in Packaged VLSI Chips

Virginia Martín Hériz[1,2], Je-Hyoung Park[1], Travis Kemper[1]
Sung-Mo Kang[1] and Ali Shakouri[1,3]
[1] Baskin School of Engineering, University of California Santa Cruz.
[2] Ecole Nationale Supérieure des Télécommunications, Paris.
[3] ali@soe.ucsc.edu

*Abstract*- **Thermal aware routing and placement algorithms are important in industry. Currently, there are reasonably fast Green's function based algorithms that calculate the temperature distribution in a chip made from a stack of different materials. However, the layers are all assumed to have the same size, thus neglecting the important fact that the thermal mounts which are placed underneath the chip can be significantly larger than the chip itself. In an earlier publication, we showed that the image blurring technique can be used to calculate quickly temperature distribution in realistic packages. For this method to be effective, temperature distribution for several point heat sources at the center and at the corner and edges of the chip should be calculated using finite element analysis (FEA) or measured. In addition, more accurate results require correction by a weighting function that will need several FEA simulations. In this paper, we introduce the convolution by images that take the symmetry of the thermal boundary conditions into account. Thus with only "two" finite element simulations, the steady-state temperature distribution for an arbitrary complex power dissipation profile in a packaged chip can be calculated. Several simulation results are presented. It is shown that the power blurring technique together with the method of images can reproduce the temperature profile with an error less than 0.5%.**

## I. INTRODUCTION

The traditional approach in order to calculate the temperature distribution in a given solid involves solving the heat equation with the appropriate boundary conditions. The most common techniques for solving a generic PDE are finite differences and finite elements, which are usually performed in the time or in the frequency domain. However, their accuracy comes at the price of long execution times, and exhaustive CPU and memory usage. Since the computation time rises significantly with the number of elements, this approach is unpractical if we are to integrate it in an interactive place-and-route IC design program. Besides, FEA programs require a thorough design of the meshing to attain convergence, a procedure which cannot be easily automated for a complex geometry or loads.

A simplification of the problem may provide us with simpler tools to deal with it. Let us examine the geometry of a typical IC: the chip is mounted over successive layers of highly heat conducting material, copper for the most cases.

Whereas the components are fabricated on a silicon die several hundreds of microns thick, the depth reached by the diffusions rarely goes over 1.5μm. Therefore, for practical purposes, we can consider these heat sources to be located on the surface, and restrict our calculations to the top surface of the silicon die, instead of dealing with the whole tridimensional problem.

Over the past few years several groups have developed a number of strategies to tackle this 2D problem, based mainly on the application of Green's functions. The earliest approach of this kind was offered by Cheng et al. [1] by modeling the chip as a semi infinite solid. The heat sources were treated as points internal to the silicon for ease of calculation. The closed form of the Green function for this system is straightforward and was approximated by a piecewise linear function. The contribution of each source was then added up to yield the total temperature distribution. In spite of being a very simplistic approach, it managed to locate hot spots with a remarkable degree of accuracy.

The next step in complexity consists of considering the finite dimensions of the thermal packaging, as well as the different materials it is made of. The Green function for infinite multilayered structures has been calculated analytically by several groups and is readily available online for free. A version for the bounded box can be calculated as an infinite linear combination of these expressions by mirroring. Zhan et al. [2,3] have devised a fast algorithm based on such analytical Green function where the associated convolution is performed in the frequency domain, resulting in one of the fastest procedures for calculating temperature profiles available at present.

However, what most groups have failed to do so far is dealing with the real geometry of the heat spreaders underneath the chip. All the previous literature assumes they have the same dimensions as the silicon die, while they are in fact, arranged in layers where the one on top has a smaller surface area than the one immediately underneath, giving it the looks of a pyramid. This simplification enables the use of analytic solutions: while Green functions for a multilayered box are common in the literature of the field, a version for a pyramidal structure has never been calculated, or even shown to exist in a closed form. Nevertheless this assumption is not realistic: the fact that heat sinks are far larger than the silicon bulks on top, has a dramatic effect in temperature: a typical copper mount of 7cm×7cm under a 1cm×1cm silicon die can reduce the temperature by several tens of degrees with respect to a mount the same size of the die. Besides, the shape of the temperature profile is also affected significantly by the size of the thermal mount. Whereas the first effect can be circumvented by an appropriate scaling of the convection coefficient, and thus reducing the average temperature rise, the second is not that straightforward.

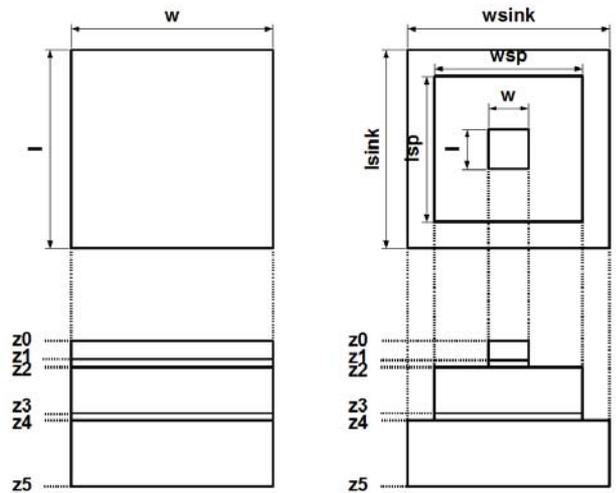

Figure 2: "Dice" and "Pyramid" geometries.

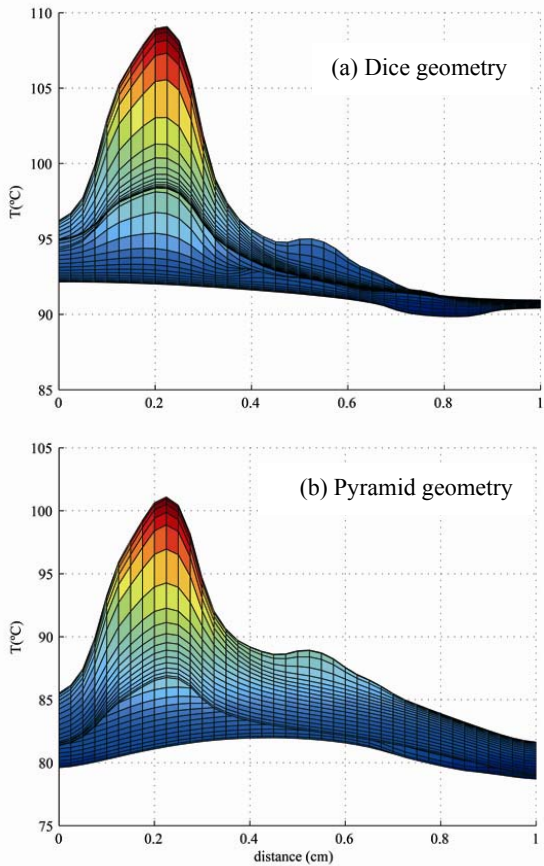

Figure 1: Temperature profiles created by the "µ-processor" power distribution in figure 3(c). The shape as well as the average temperature are affected substantially by the geometry of the thermal mount.

Figure 1 illustrates the temperature distribution resulting from applying a given power distribution to two chips with different thermal mounts: 1(a) was mounted over two copper bulks the same size as the silicon die, while in 1(b) the copper heat spreaders were 10 and 50 times larger in surface than the chip. Separating the three layers (silicon, spreader and sink) are two thermal interfaces, making a total of five layers. These geometries are displayed in Figure 2, and are labeled dice and pyramid, respectively. In order to obtain a realistic temperature profile in the first chip, the convection coefficient was scaled accordingly, so that total convection remains the same for both chips. Note that even in this case, the average temperature rise will still be slightly different, since the mounts are different in shape.

At the time of the writing of this paper Bagnoli et al. [4] have managed to find an analytical system of equations that relates the temperature and heat flux at the material interfaces of the pyramid structure. These equations can be discretized and solved by the usual matrix inversion. While this technique can be applied to the pyramid multilayer structure, it remains ineffective in terms of time and resources, even though the calculations remain essentially 2D.

The main drawback of tackling the more realistic pyramidal geometry is that simple analytic expressions no longer exist, and we are resorted to use semi analytical methods where a part of the algorithm must rely either on empirical parameters or previous simulations if we want to be reasonably fast.

Kemper et al. [5] have already made contributions in this area of research, with a scheme deemed power blurring: The surface under study is divided into several regions and a sample response to the heat impulse is evaluated on each of them. The thermal results are obtained by direct convolution of the power distribution and the impulse response in each area.

In this paper we present a comparative study of two convolution-based fast strategies: the first one is a refinement of the previous power blurring scheme, while the second one takes advantage of the method of images used in electrostatics. Hereon, we will refer to these methods as Convolution by Regions (CR), and Convolution by Images (CI) respectively. While CR may be applied to pyramid geometries directly, the convolution by images needs further corrections. Two different strategies were devised, deemed Convolution by Images Type I (CI1) and Type II (CI2).

All three methods (CR, CI1 and CI2) show a significant improvement on the error rate over the original power blurring scheme, while taking into account the real shape of the thermal mounts.

This paper will be organized as follows: section 2 will provide an overview of these methods, section 3 will show the numerical simulations linked to the two geometries mentioned above (Figure 2) and, finally, the conclusion will summaries the key findings.

## 2. OVERVIEW OF THE PROPOSED METHODS

The top surface of the circuit under study is divided into a regular mesh of N×M elements. For each element we will retrieve the temperature at the center. Therefore we will end up with a N×M matrix where each point represents the temperature at a given location of the chip.

As outlined in the introduction, our three strategies are all based on the convolution of the original power with a certain impulse response h. Instead of generating such a function analytically, we have resorted to the popular FEA program ANSYS [6]. Fortunately, we only need to obtain the impulse heat response once for each mount, so after it is computed it may be stored and used for further calculations.

TABLE I
Parameter values

| Name | Value (cm) | Name | Value (cm) |
|---|---|---|---|
| w | 1 | $l_{die}$ | 1 |
| $w_{sink}$ | 7 | $z_0$ | 0 |
| $w_{spreader}$ | 3 | $z_1$ | 0.05 |
| $w_{die}$ | 1 | $z_2$ | 0.07 |
| l | 1 | $z_3$ | 0.22 |
| $l_{sink}$ | 7 | $z_4$ | 0.24 |
| $l_{spreader}$ | 3 | $z_5$ | 0.74 |

TABLE II
Material property values used in the ANSYS simulations

| Material | Thermal conductivity (W/cm·K) | Density (Kg/cm³) | Specific heat (J·kg/K) |
|---|---|---|---|
| Si | 1.25 | 0.00233 | 700 |
| Cu | 3.95 | 0.00893 | 397 |
| Solder | 0.3 | 0.00193 | 15 |

Figure 2 shows the two structures we used to obtain the unit response: the left figure depicts a regular stack of materials, which is the thermal mount assumed by all the fast algorithms available in the literature, and the figure on the left shows its more realistic, yet less symmetric, counterpart. We are referring to these geometries as dice and pyramid, respectively. Unless otherwise noted, the dice mount is used to generate the impulse response. In order to achieve this, we make the size of the point heat source as small as possible compared to the size of the package, so that it is close to a delta function approximation.

Note that the meshing we use in our ANSYS simulations is not related to the N×M divisions we have outlined in our chip. These divisions are only intended for the matrix multiplication algorithms we have devised and they are evenly spaced for simplicity purposes, whereas the ANSYS meshing has had to be adapted to each power distribution in order to achieve convergent independent of the mesh size.

### 2. 1. Convolution by Regions

The simplest convolution approach would be to take the impulse response at the center of the chip and convolute it with the power distribution on the chip. While this would produce realistic results in the center, the error figures would be significantly high due to the border effects. There the impulse response near the edges behaves differently that at the center, this is the key insight of the CR method.

Instead of using the same response for the whole chip, we divided it into regions and used a different response for each region. For our particular case, we divided it into three regions: the corners, the edges, and the middle. As this approach still yielded noticeable error densities in the boundary between regions, we have modified it slightly so that now there is a transition region. There we use an impulse response which is an interpolation of those of the neighboring regions. The algorithm has been implemented so that there is no additional computational load with respect to the previous, simpler, version: we only needed to calculate three N×M multiplications and convolutions, plus a summation. With respect to the FEA simulations, we only need to perform three for each thermal mount.

2. 2. Convolution by Images

This method can be divided into three steps:

- Calculation of the unit heat impulse response at the center of the chip.
- Computation of the temperature profile taking into account that the chip size is finite and that there are edge effects.
- Corrections for the pyramid geometry.

The temperature point response function for a bounded rectangular domain $h_{BOX}$ can easily be calculated as an infinite summation of displaced and mirrored versions of h.

$$h_{BOX} = \sum_n \sum_m [h(x-nL-x_0, y-mL-y_0) \\ + h(x-nL-x_0, y-mL+y_0) \\ + h(x-nL+x_0, y-mL-y_0) \\ + h(x-nL+x_0, y-mL+y_0)]$$

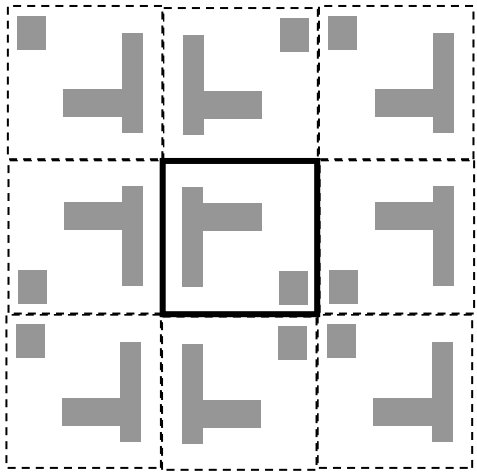

Figure 3: showing an arbitrary heat distribution on the chip and the eight nearest neighbor mirror images of this distribution.

This is because the adiabatic thermal boundary condition at e.g. one edge of the rectangular domain can be achieved by placing two sources mirror image of each other on the two sides of the boundary (see Figure 3). Since the impulse response is a bell shaped function that tends to a constant far away from the source, most of the multiple image terms from parallel surfaces can be neglected, save for the nearmost neighbors:

$$h_{BOX} \cong h(x-x_0, y-y_0) + h(x-x_0, y+y_0) + h(x-x_0, y-W-y_0) \\ + h(x+x_0, y-y_0) + h(x+x_0, y+y_0) + h(x+x_0, y-W-y_0) \\ + h(x-L-x_0, y-y_0) + h(x-L-x_0, y+y_0) + h(x-L-x_0, y-W-y_0)$$

Since the impulse response depends on the spatial coordinates, the resulting temperature distribution cannot be calculated by direct convolution, at least in principle. However, by working on the equations in the frequency domain, the mirroring of the images is mathematically equivalent to the mirroring of the fixed power distribution with respect to different boundaries:

$$T(x,y) = h(x,y) * [P(x,y) + P(x,-y) + P(x, y-W) \\ + P(-x, y) + P(-x,-y) + P(-x, y-W) \\ + P(x-L, y) + P(x-L,-y) + P(x-L, y-W)]$$

2. 3. Corrections for the pyramid geometries

2. 3. 1. Convolution by Images Type I

While the previous solution is remarkably accurate for dice geometries (average relative error below 0.4% with respect to FEA simulations), we already outlined in the introduction that it is not necessarily the case for pyramid structures. From our finite element numerical simulations we have concluded that the convolution by images gives reasonable relative error rates near the edges, whereas near the center a simple convolution is more accurate. Such is the effect of a large thermal mount: even though the silicon die is much smaller, the impulse response behaves as if it were much larger, and so the effect of the images is only important in the edges. This realization has encouraged us to combine both strategies to obtain the temperature distribution in the pyramid geometry: we will divide the die into two regions, a square in the middle, where we will apply convolution, and the corners and edges, where we will use images. There will also be a transition region where we will apply a linear combination of the two solutions. We have based the size of these regions on the simulations with a uniform power distribution, since it places all locations within the chip in an equal footing. We applied the method for different sizes of the center panel, as well as the interpolation region and determined the optimum. For a 41×41 element grid, it resulted in a 37×37 element middle region, and a interpolation zone 2 element wide.

On the other hand, we can no longer estimate the overall temperature rise in the same way we did in the dice case. This is due to the fact that, unlike dice geometries, the impulse response in a pyramid geometry does not tend to a constant at far away distances, but rather it decreases steadily even far away from the source, so there will be a certain truncation error. Nevertheless, we can still estimate the overall temperature rise by computing the average temperature of the surface by alternative means. Since the temperature profile created by a unit heat source is not uniform, it is reasonable to expect the average temperature rise of the surface to depend on its location. We can thus measure the average temperature rise created by a heat impulse as a function of position (x, y). This function depends undoubtedly on the geometry of the thermal mount underneath, so it has to be measured for every device. Afterwards, it may be used to compute the average temperature caused on the chip by any power distribution.

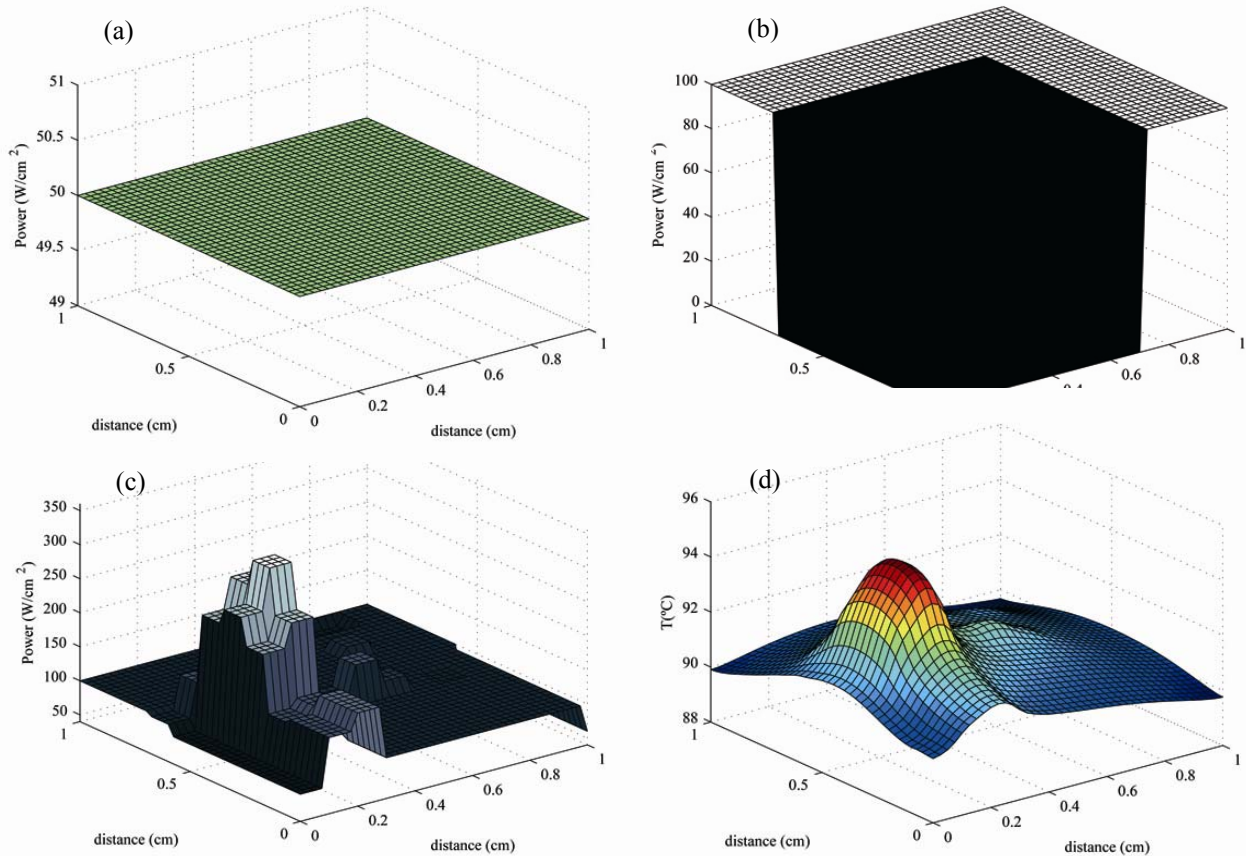

Figure 4: The three power distributions selected for our study, and a sample temperature distribution calculated by ANSYS. (a) "Uniform" power distribution. (b) "Edge" power distribution. (c) "Microprocessor" power distribution. (d) Temperature profile calculated for (c).

2. 3. 2. Convolution by images Type II

The starting point is, like in the previous strategy, the convolution by images. The relative error between the real temperature distribution and that calculated by images, for a uniform power distribution, was calculated:

$$e_r = \frac{T_{images} - T_{real}}{T_{real}}$$

$e_r$ was then used as position-dependent scaling factor for the relative deviation between the real temperature and the temperature found by the method of images, for any kind of power distribution.

Extensive simulations have proved this method to be the most effective when calculating temperature distributions in pyramid geometries, as the next section will show. Besides, it is the most economic in terms of FEA simulations, since only two are needed for a particular geometry: one to construct the impulse response, and another to calculate the temperature under a uniform power load, which will in turn be used to derive $e_r$ as described by the previous equation.

TABLE III
Error rates with respect to ANSYS for the different methods and geometries discussed.

**Power Distribution: "Uniform"**

| *Dice* Geometry | | | | *Pyramid* Geometry | | | |
|---|---|---|---|---|---|---|---|
| method | Avg. Err. | Max. Err. | Hot spot | method | Avg. Err. | Max. Err. | Hot spot |
| CR | 0.56% | 4.45% | 4.45% | CR | 0.70% | 5.98% | 0.18% |
| CI | 0.01% | 0.01% | 0.01% | CI1 | 0.77% | 2.70% | 0.97% |
| | | | | CI2 | NA | NA | NA |

**Power Distribution: "Edge"**

| *Dice* Geometry | | | | *Pyramid* Geometry | | | |
|---|---|---|---|---|---|---|---|
| method | Avg. Err. | Max. Err. | Hot spot | method | Avg. Err. | Max. Err. | Hot spot |
| CR | 2.15% | 10.57% | 10.57% | CR | 0.98% | 7.13% | 0.58% |
| CI | 1.83% | 5.46% | 3.10% | CI1 | 1.19% | 4.59% | 0.37% |
| | | | | CI2 | 0.48% | 1.14% | 0.27% |

**Power Distribution: "µ-processor"**

| *Dice* Geometry | | | | *Pyramid* Geometry | | | |
|---|---|---|---|---|---|---|---|
| method | Avg. Err. | Max. Err. | Hot spot | method | Avg. Err. | Max. Err. | Hot spot |
| CR | 1.15% | 10.03% | 0.24% | CR | 0.72% | 6.56% | 0.88% |
| CI | 0.39% | 2.11% | 0.32% | CI1 | 0.70% | 3.76% | 0.59% |
| | | | | CI2 | 0.27% | 1.89% | 1.06% |

TABLE IV
Comparison between the executions times of ANSYS and our three algorithms for the pyramid geometry.

| Power Dist. | CR | CI1 | CI2 | ANSYS |
|---|---|---|---|---|
| edge | 0.32s | 0.16s | 0.12s | 54.00s |
| uniform | 0.31s | 0.18s | 0.14s | 37.97s |
| µprocessor | 0.33s | 0.18s | 0.10s | 56.24s |

3. RESULTS

As outlined previously, Figure 2 depicts the geometries used in our simulations. The parameters are detailed in Table 2. Table 1 provides a summary of the material properties. The heat source was calculated as the temperature distribution produced by a power density of $1W/cm^3$ applied over an area of $1/41 \times 41 cm^2$. It was then sampled at 1/41cm-long intervals, in both x and y directions.

The simulations on the dice geometry are provided as a control. The symmetry of such geometries makes the sources of error more tractable than in the pyramid case, and provide us with a rough estimate of the errors we could expect by applying our methods. In the case of the convolution by images, the error rates are due to quantization, the truncation of the infinite images series and the fact that the impulse response for an infinite domain in x and y was approximated by one in a rectangle. Concerning the Convolution by Regions scheme, it is more difficult to pinpoint a localized cause of error as in the convolution by images, since it is by its very nature, a very approximate approach and it depends of the choice of boundaries between different regions.

However, the fact that we are approximating the impulse response in an infinite domain by that in a bounded domain proves specially damaging in dice geometries, and therefore

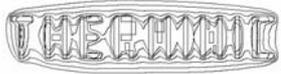
*Budapest, Hungary, 17-19 September 2007*

we will see larger error figures than in their pyramid counterparts. A possible solution to this problem is to use an impulse response evaluated in a region larger than the chip itself. Simulations with such impulse responses were performed in dice geometries and produced error rates close to those in pyramid geometries. Unfortunately we can only use this approach in CI. In CR we are evaluating the impulse response in the corners and edges, and thus breaking the symmetry we rely on to extend the domain used in the evaluation of the impulse response. For this reason the simulations presented in this paper make use of impulse responses covering exactly the area of the chip, being the largest domain applicable to all methods.

Figure 4 depicts three selected power distributions we used in our tests. 4(a), being constant, was aimed at revealing the weakest regions for each algorithm. 4(b) was intended to provide a worst case scenario by concentrating all the power on the edges, since we know that CR is especially sensitive to them. 4(c) is a realistic representation of what a power distribution on a modern-day ASIC might look like and the temperature profile it creates on the surface of the chip is displayed in 3(d) (as simulated by ANSYS).

Table 3 shows the average, the maximum and the error in the hottest spot for the three power maps under discussion in both geometries. The errors are calculated as relative to simulations performed by ANSYS. On the other hand, Figures 5(a), 5(b) and 5(c) feature the cross section of the temperature profiles calculated with our algorithms and simulated by ANSYS.

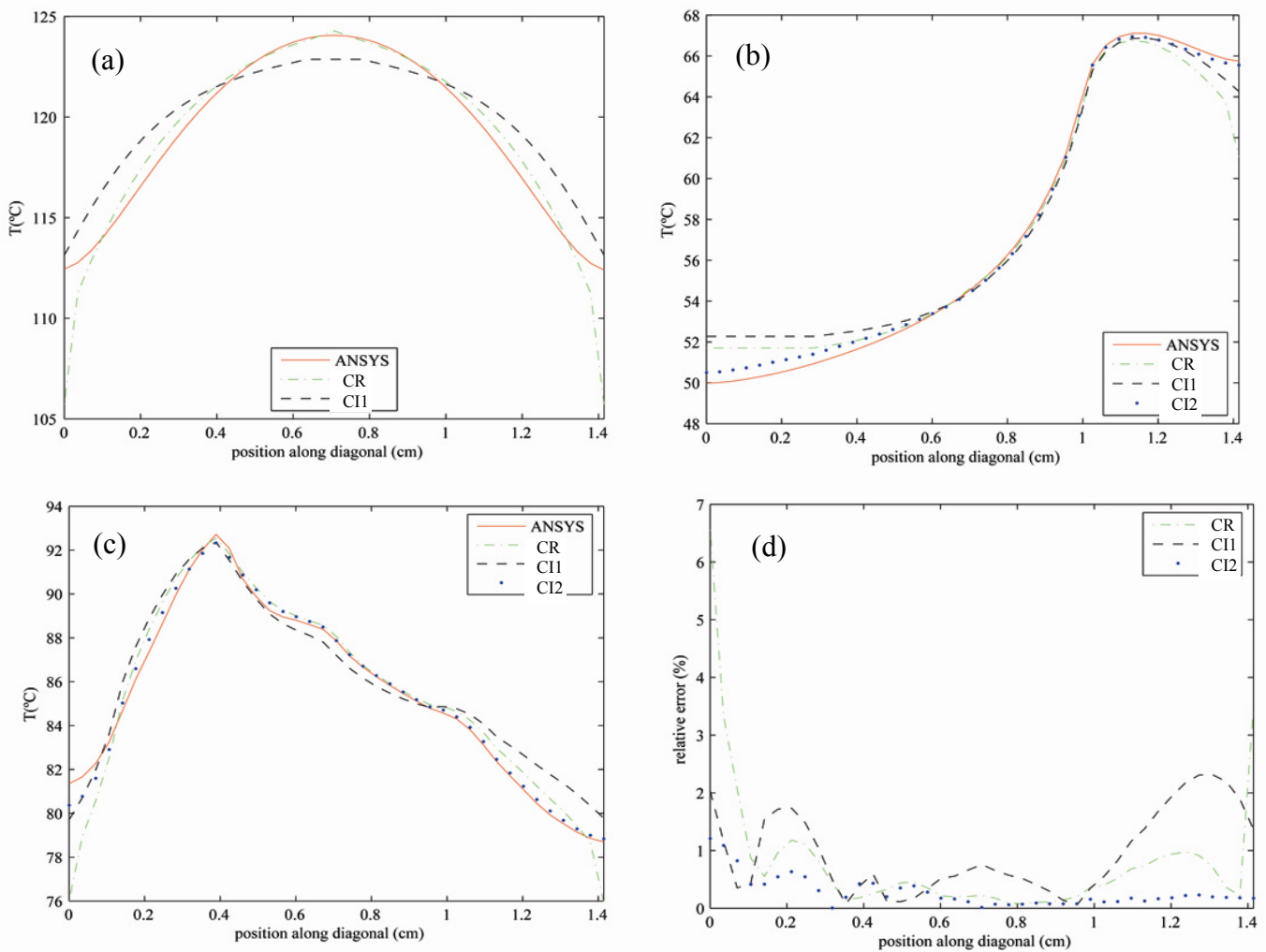

Figure 5: Diagonal section of the temperature and error profiles for different power distributions and algorithms: (a) "uniform" power distribution. (b) "edge" power distribution. (c) "μ-processor" temperature distribution. (d) relative error of the "μ-processor" distribution

By looking at the figures, we can see that the convolution by images performs better than CR in dice geometries, yielding average errors no larger than 2% and maximum errors around 5% for the worst cases. This does no seem surprising, given that the error is due to truncation of the impulse response, whereas CR is a more approximate method.

As for the pyramid geometries, CR keeps the average error below 1%. The maxima, however, are a little less optimistic, amounting up to 7%, since this approach fails to simulate the adiabatic boundary condition properly. The border effects are accurately predicted by any of the Image-based type schemes, as it can be appreciated in Figures 5(a), 5(b) and 5(c). Even though CI1 has slightly higher average error rates than CR, the maxima are halved.

In any case, the best results are provided by CI2, with maxima under 2% and averages no larger than 0.5%, making it between two and five times more accurate than the other two approaches. The fact that CI2 has proved to be the best algorithm becomes evident in Figure 5(d), which illustrates a cross section of the error profile produced by all three methods under discussion, again with respect to ANSYS simulations.

The execution times of our algorithms against those of ANSYS are displayed in Table 4. Our convolution methods take about the same time to run, and it is clear that that they outperform ANSYS by approximately two orders of magnitude. Besides, execution time in ANSYS is heavily determined by the complexity of the power distribution, which determines the number of elements in the mesh necessary for convergence. Our methods, on the contrary, are independent of the input power distribution: since our impulse response functions have been obtained by sampling of the real heat response, our algorithms converge regardless of the size of the grid. The execution times are thus constant for a particular grid size, no matter how complex the power distribution might be.

4. CONCLUSION

This paper presents three alternative ways to compute the temperature profiles in modern day ICs. Unlike any other method available in the literature, they are capable of dealing with realistic package geometries, while keeping the execution times to the order of fractions of a second and, for our best algorithm, CI2, the average error rates as low as 0.5%.

In modern day applications, the heat spreader and sink are known well before designing the IC on top, and are rarely changed afterwards. As a result, the temperature point response function of our algorithms that can be related to the package geometry only need to be calculated once, and therefore are not crucial as far as speed is concerned. The key issues are the calculations involving the distribution of heat sources, which have to be executed every time the components are rearranged.

The three schemes under discussion rely on previous FEA simulations for the unit impulse response and the additional correction functions needed for pyramid geometries. On the light of this paper, such calculations can be done offline. On the contrary, the temperature calculations have to be performed for every arrangement of the electronic components. As discussed, this can be done in real time, since it only takes a few multiplications and FFTs each time.

Further improvements can remove the need for ANSYS, making our algorithms fully analytical. The unit heat response may be obtained from the analytical expression for the Green function for a dice geometry as calculated by [3], while the correction functions may be built by solving the system of equations outlined by [4]. In such a fashion, our algorithms may be integrated in commercial softwares as a module to assist in thermal-aware placement and routing. Note that the procedures we have described to obtain the temperature distributions remain essentially the same, independently of the method (ANSYS or analytical) we use to calculate the impulse response and the corrections.